\newif\ifpdf
\ifx\pdfoutput\undefined
\pdffalse 
\else
\pdfoutput=1 
\pdftrue \fi

\newif\iffinal
\finalfalse	
\finaltrue 

\documentclass[reqno,twoside,11pt]{amsart}
\iffinal\else\usepackage[notref,notcite]{showkeys}\fi
\iffinal\else\IfFileExists{pdfsync.sty}{\usepackage{pdfsync}}{}\fi
\usepackage{cite}
\usepackage{amsmath}
\usepackage{amsfonts}
\usepackage{amssymb}
\usepackage{epsfig}
\usepackage{verbatim}

\IfFileExists{epsf.def}{\input epsf.def}{\usepackage{epsf}}

\usepackage{mathptmx}
\usepackage{mathrsfs}
\DeclareFontFamily{OT1}{eusb}{} \DeclareFontShape{OT1}{eusb}{m}{n} {<5> <6> <7> <8> <9> <10> <11> <12> <14.4> eusb10}{}
\DeclareMathAlphabet{\eusb}{OT1}{eusb}{m}{n}

\DeclareFontFamily{OT1}{eusm}{} \DeclareFontShape{OT1}{eusm}{m}{n} {<5> <6> <7> <8> <9> <10> <11> <12> <14.4> eusm10}{}
\DeclareMathAlphabet{\eusm}{OT1}{eusm}{m}{n}

\DeclareFontFamily{OT1}{eufm}{} \DeclareFontShape{OT1}{eufm}{m}{n} {<5> <6> <7> <8> <9> <10> <11> <12> <14.4> eufm10}{}
\DeclareMathAlphabet{\mathfrak}{OT1}{eufm}{m}{n}

\DeclareFontFamily{OT1}{fraktura}{}
\DeclareFontShape{OT1}{fraktura}{m}{n} {<5> <6> <7> <8> <9> <10> <11> <12> <13> <14.4> [1.1] eufm10}{}
\DeclareMathAlphabet{\fraktura}{OT1}{fraktura}{m}{n}

\DeclareFontFamily{OT1}{cmfi}{} \DeclareFontShape{OT1}{cmfi}{m}{n} {<5> <6> <7> <8> <9> <10> <11> <12> <13> <14.4> [0.9] cmfi10}{}
\DeclareMathAlphabet{\cmfi}{OT1}{cmfi}{b}{n}

\DeclareFontFamily{OT1}{cmss}{} \DeclareFontShape{OT1}{cmss}{m}{n} {<5> <6> <7> <8> <9> <10> <11> <12> <13> <14.4> cmss10}{}
\DeclareMathAlphabet{\cmss}{OT1}{cmss}{m}{n}

\newtheoremstyle{thm}{1.5ex}{1.5ex}{\itshape\rmfamily}{} {\bfseries\rmfamily}{}{2ex}{}

\newtheoremstyle{def}{1.5ex}{1.5ex}{\rmfamily\sl}{} {\bfseries\rmfamily}{}{2ex}{}

\newtheoremstyle{rem}{1.3ex}{1.3ex}{\rmfamily}{} {\bfseries\rmfamily}{}{2ex}{}

\newtheoremstyle{ass}{1.5ex}{1.5ex}{\rmfamily\sl}{} {\bfseries\rmfamily}{}{2ex}{}

\newenvironment{proofsect}[1] {\vskip0.1cm\noindent{\rmfamily\itshape#1.}}{\qed\vspace{0.15cm}}

\theoremstyle{thm}
\newtheorem{theorem}{Theorem}[section]
\newtheorem{lemma}[theorem]{Lemma}
\newtheorem{proposition}[theorem]{Proposition}

\newtheorem*{Main Theorem}{Main Theorem.}
\newtheorem{corollary}[theorem]{Corollary}
\newtheorem{assumption}[theorem]{Assumptions}

\theoremstyle{def}

\theoremstyle{rem}
\newtheorem{remark}[theorem]{{Remark}}
\newtheorem{remarks}[theorem]{{Remarks}}

\numberwithin{equation}{section}

\renewcommand{\section}{\secdef\sct\sect}
\newcommand{\sct}[2][default]{\refstepcounter{section}
\addcontentsline{toc}{section}
{{\tocsection {}{\thesection}{\!\!\!\!#1\dotfill}}{}}
\vspace{0.7cm}
\centerline{ 
\scshape\arabic{section}.\ #1} \nopagebreak \vspace{0.2cm}}
\newcommand{\sect}[1]{
\vspace{0.4cm} \centerline{\large\scshape\rmfamily #1}
\vspace{0.2cm}}

\renewcommand{\subsection}{\secdef\subsct\sbsect}
\newcommand{\subsct}[2][default]{\refstepcounter{subsection}
\addcontentsline{toc}{subsection}
{{\tocsection{\!\!}{\hspace{1.2em}\thesubsection}{\!\!\!\!#1\dotfill}}{}}
\nopagebreak\vspace{0.45\baselineskip} {\flushleft\bf
\arabic{section}.\arabic{subsection}~\bf #1.~}
\\*[3mm]\noindent
\nopagebreak}
\newcommand{\sbsect}[1]{\vspace{0.1cm}\noindent
\textbf{#1.~}\vspace{0.1cm}}

\renewcommand{\subsubsection}{%
\secdef \subsubsect\sbsbsect}
\newcommand{\subsubsect}[2][default]{%
\refstepcounter{subsubsection}
\addcontentsline{toc}{subsubsection}{{\tocsection{\!\!}
{\hspace{3.05em}\thesubsubsection}{\!\!\!\!#1\dotfill}}{}}
\nopagebreak
\vspace{0.15\baselineskip} \nopagebreak {\flushleft\rmfamily
\itshape\arabic{section}.\arabic{subsection}.\arabic{subsubsection}
\ \rmfamily #1\/.}\ }
\newcommand{\sbsbsect}[1]{\vspace{0.1cm}\noindent
\rmfamily \itshape
\arabic{section}.\arabic{subsection}.\arabic{subsubsection} \
\sffamily #1\/.\ }

\iffinal

\else

\fi

\renewcommand{\caption}[1]{%
\vglue0.5cm
\refstepcounter{figure}
\begin{minipage}{0.9\textwidth}\small {\sc Figure~\thefigure. }#1\end{minipage}}

\newcommand{\dist}{\operatorname{dist}}

\newcommand{\textd}{\text{\rm d}\mkern0.5mu}

\newcommand{\texte}{\text{\rm  e}\mkern0.7mu}

\newcommand{\Cov}{\text{\rm \Cov}}

\newcommand{\EE}{\mathcal E}

\newcommand{\II}{\mathcal I}

\newcommand{\PP}{\mathcal P}

\newcommand{\D}{\mathbb D}

\newcommand{\N}{\mathbb N}

\newcommand{\R}{\mathbb R}
\newcommand{\BbbS}{\mathbb S}

\newcommand{\Z}{\mathbb Z}

\newcommand{\scrH}{\mathscr{H}}

\newcommand{\cc}{{\text{\rm c}}}

\def\myffrac#1#2 in #3{\raise 2.6pt\hbox{$#3 #1$}\mkern-1.5mu\raise 0.8pt\hbox{$#3/$}\mkern-1.1mu\lower 1.5pt\hbox{$#3 #2$}}
\newcommand{\ffrac}[2]{\mathchoice%
	{\myffrac{#1}{#2} in \scriptstyle}
	{\myffrac{#1}{#2} in \scriptstyle}
	{\myffrac{#1}{#2} in \scriptscriptstyle}
	{\myffrac{#1}{#2} in \scriptscriptstyle}
}

\newcommand{\hate}{\hat{\texte}}

\newcommand{\SO}{\text{\rm SO}}
\newcommand{\justO}{\text{\rm O}}

\newcommand\sign{\text{\rm sign}}

\title[Absence of magnetism in continuous-spin systems]
{\large Absence of magnetism in continuous-spin systems\\with long-range antialigning forces}
\author[M.~Biskup and N. Crawford]{Marek Biskup$^{1,2}$\, and\, Nicholas Crawford$^3$}

\begin{document}
\thanks{\hglue-4.5mm\fontsize{9.6}{9.6}\selectfont\copyright\,2011 M.~Biskup and N. Crawford. Reproduction, by any means, of the entire
article for non-commercial purposes is permitted without charge.\vspace{2mm}}
\maketitle

\vspace{-5mm}
\centerline{${}^1$\textit{Department of Mathematics, UCLA, Los Angeles, California, U.S.A.}}
\centerline{${}^2$\textit{School of Economics, University of South Bohemia, \v Cesk\'e Bud\v ejovice, Czech Republic}}
\centerline{${}^3$\textit{Department of Industrial Engineering, Technion, Haifa, Israel}}

\begin{abstract}
We consider continuous-spin models on the $d$-dimensional hypercubic lattice with the spins~$\sigma_x$ \emph{a priori} uniformly distributed over the unit sphere in~$\R^n$ (with~$n\ge2$) and the interaction energy having two parts: a short-range part, represented by a potential~$\Phi$, and a long-range antiferromagnetic part $\lambda|x-y|^{-s}\sigma_x\cdot\sigma_y$ for some exponent~$s>d$ and $\lambda\ge0$. We assume that~$\Phi$ is twice continuously differentiable, finite range and invariant under rigid rotations of all spins. For $d\ge1$,~$s\in(d,d+2]$ and any $\lambda>0$, we then show that the expectation of each~$\sigma_x$ vanishes in all translation-invariant Gibbs states. In particular, the spontaneous magnetization is zero and block-spin averages vanish in all (translation invariant or not) Gibbs states. This contrasts the situation of~$\lambda=0$ where the ferromagnetic nearest-neighbor systems in~$d\ge3$ exhibit strong magnetic order at sufficiently low temperatures. Our theorem extends an earlier result of A.~van Enter ruling out magnetized states with uniformly positive two-point correlation functions.
\end{abstract}

\section{Introduction}
\label{sec1}\noindent
In the last couple of years, there has been renewed interest by mathematicians in the behavior of lattice models with spins interacting via long-range (e.g., dipole-dipole) interactions. This has partially been motivated by advances in quasi-two-dimen\-sional physics, but much of it derives from the theoretical challenge that these systems seem to pose to existing methods of proof. Indeed, long-range interactions are generally quite hard to handle and most of the techniques that control nearest-neighbor systems are of little or no use when short-range and long-range forces are mixed together in competition.

For definiteness of further discussion, let us consider a system of $\justO(n)$-spins $\sigma_x$, with $x\in\Z^d$ and each $\sigma_x$ being \emph{a priori} uniformly distributed over the unit sphere in~$\R^n$. The interaction between the spins is described by the Hamiltonian
\begin{equation}
\label{1.1}
\scrH(\sigma):=-J\sum_{\langle x,y\rangle}\sigma_x\cdot\sigma_y+\sum_{x\ne y}\frac\lambda{|x-y|^s}\sigma_x\cdot\sigma_y.
\end{equation}
Here the first sum goes over pairs of nearest neighbors in~$\Z^d$, the long-range coupling strength obeys~$\lambda\ge0$ and the interaction is summable by the assumption~$s>d$. The equation \eqref{1.1} defines the model with \emph{scalar} long-range interaction; to get the \emph{dipole} model one needs to change the second summand into $\sum_{i,j=1}^dK_{xy}(i,j)\sigma_x^i\sigma_y^j$ where $K_{xy}(i,j)=-\partial_{x_i}\partial_{x_j}\varphi(x-y)$ for $\varphi$ denoting the Coulomb potential. An intriguing feature of the dipole model is that the sign and strength of the interaction depend sensitively on the orientation of the spins with respect to the line segment connecting their spatial positions.

A key question concerning the model \eqref{1.1} is the existence of \emph{stripe states}, i.e., Gibbs measures supported on configurations with alternating stripes of spins oriented in different directions. For certain 1D and 2D systems, existence of such states has been established mathematically in papers by Giuliani, Lebowitz and Lieb~\cite{GLL06,GLL07,GLL09}, albeit only at zero temperature. Currently there seem to be no rigorous results concerning the stripe order at positive temperatures, perhaps with the exception of the work on the Kac limit of the free energy in certain spin systems with modulated interactions of indefinite sign, cf~Gates and Penrose~\cite{Gates-Penrose}, Pisani, Smith and Thompson~\cite{Pisani-Smith-Thompson} and Pisani and Thompson~\cite{Pisani-Thompson}. For the dipole-dipole system, Giuliani~\cite{Giuliani} recently completed an argument (building on Fr\"ohlich, Simon and Spencer~\cite{FSS} and Fr\"ohlich and Spencer~\cite{Frohlich-Spencer}) that establishes the existence of an \emph{orientational} long-range order --- although only for the situation without the nearest-neighbor term.

The aim of this paper is to resolve a simpler question: the existence/absence of \emph{magnetic} (or, more precisely, \emph{ferromagnetic}) order. Our principal result is that, for the model in \eqref{1.1} with~$n\ge2$ and exponents $s\in(d,d+2]$, as soon as~$\lambda>0$, the expectation of $\sigma_x$ vanishes in all translation-invariant states at all positive temperatures. A consequence of this is that the spontaneous magnetization --- defined by the derivative of the pressure with respect to the external field --- vanishes as well, and so do the block-spin averages in all (translation-invariant or not) Gibbs states. This is somewhat surprising because when~$\lambda=0$ and~$d\ge3$ (and~$J>0$) the system \eqref{1.1} shows a strong magnetic order at low temperatures (Fr\"ohlich, Simon and Spencer~\cite{FSS}). Our theorem provides novel information even in dimensions~$d=1$ and~$2$ because the Mermin-Wagner theorem does not apply to the whole range of exponents~$s$ we wish to consider; cf Remark~\ref{R}(4).

The problem of magnetic order in model \eqref{1.1} has quite a long history. To our knowledge, it first appears in studies by van Enter~\cite{vanEnter,vanEnterPRB} on the ``instability'' of phase diagrams (and validity of the Gibbs-phase rule) under ``irrelevant'' perturbations. Specifically, in~\cite{vanEnterPRB} it was shown that certain natural magnetically-ordered states in short-range ferromagnetic spin systems are destabilized --- in the sense of failing to minimize the Gibbs variational problem --- by adding the above long-range antiferromagnetic interaction with exponents $d<s<d+2$. A subtle point is that the assumption made in \cite{vanEnterPRB} on the purported magnetized state $\mu$ is that of \emph{clustering}; explicitly,
\begin{equation}
\label{E:cluster}
E_\mu(\sigma_0\cdot\sigma_x)-E_\mu(\sigma_0)\cdot E_\mu(\sigma_x)\underset{|x|\to\infty}\longrightarrow0.
\end{equation}
Along with the (natural) assumptions of translation invariance and non-vanishing value of $E_\mu(\sigma_x)$, this permits one to assume a uniform positive lower bound on $E_\mu(\sigma_x\cdot\sigma_y)$ for any $x$ and~$y$ that are sufficiently far apart.

The result of \cite{vanEnterPRB} thus rules out  the ``standard'' magnetic ordering seen in the purely ferromagnetic system, which is marked by uniform positivity of $E_\mu(\sigma_x\cdot\sigma_y)$. However, there are other ways that the system can show magnetic order without $E_\mu(\sigma_x\cdot\sigma_y)$ having a definite sign. For instance, if a typical configuration in such a state exhibits a modulated ``stripe'' structure --- with $E_\mu(\sigma_x)$ gradually turning around the ``clock'' as $x$ slides along one of the coordinate directions --- then (assuming clustering) $E_\mu(\sigma_x\cdot\sigma_y)$ will oscillate between positive and negative values. Ruling out such cases along the argument of \cite{vanEnterPRB} would require making further assumptions on \emph{how exactly} $E_\mu(\sigma_x\cdot\sigma_y)$ changes as~$y$ moves away from~$x$. Anyway, this would still not exclude the possibility of other structures --- e.g., the \emph{bubble} states or aperiodically modulated states.

Our approach overcomes these difficulties by working solely under the assumption of \emph{ergodicity} with respect to spatial translations. Conceptually, we build on an earlier paper by Biskup, Chayes and Kivelson~\cite{BCK} showing that no magnetic order exists (at any temperature) in the \emph{Ising}-spin version of the model once~$\lambda>0$ and~$s\in(d,d+1]$. In fact, the method of~\cite{BCK} would establish the same result also for~$\justO(n)$ spins for all~$s\le d+1$. However, as is also shown in~\cite{BCK}, the proof cannot extend beyond this range because the Ising-spin version of \eqref{1.1} does exhibit magnetic order at low temperatures as soon as~$s>d+1$ and~$\lambda\ll J$.
The argument of~\cite{BCK} is based on a flip of all spins in a large box and a careful accounting of the change in energy caused thereby. A key technical challenge here is to find a way to achieve the same effect via a ``continuous'' --- i.e., Mermin-Wagner like --- deformation.

The rest of this note is organized as follows: In Section~\ref{sec2} we develop the necessary foundations for the statement of our main result. In Section~\ref{sec3}, we give the main steps of the proof while deferring the technical claims to Sections~\ref{sec4} and~\ref{sec5}.

\section{Statement of the result}
\label{sec2}\noindent
Consider the $d$-dimensional hypercubic lattice~$\Z^d$, fix $n\ge2$ and let~$\BbbS^{n-1}$ denote the unit sphere in~$\R^n$. We will consider spin configurations~$\sigma:=(\sigma_x)_{x\in\Z^d}$ taking values in the product space $\Omega:=(\BbbS^{n-1})^{\Z^d}$. Let~$\tau_x$ be the shift by~$x$ on~$\Omega$, which is defined by
\begin{equation}
(\tau_x\sigma)_z:=\sigma_{x+z},\qquad z\in\Z^d.
\end{equation}
Let~$\SO(n)$ denote the group of real orthogonal $n\times n$-matrices with unit determinant. For each $\cmss R\in \SO(n)$, let $(\cmss R\sigma)_x:=\cmss R\sigma_x$ denote the global (rigid) rotation of the spin configuration~$\sigma$ by matrix~$\cmss R$. For each~$N\in\N$, consider the block
\begin{equation}
\Lambda_N:=[-N,N]^d\cap\Z^d.
\end{equation}
The definition of our model will require two objects: A function~$\Phi\colon\Omega\to\R$ representing the short-range interaction and a kernel~$(K_{xy})_{xy\in\Z^d}$ representing the coupling constants for the long-range interaction.

\begin{assumption}
\label{ass}
(1)  Suppose that there is an~$r\in\N$ such that $\Phi\colon\Omega\to\R$ depends only on $\{\sigma_x\colon x\in\Lambda_r\}$. Moreover, assume~$\sigma\mapsto\Phi(\sigma)$ is~$C^2$ (as a function on a smooth manifold) and
\begin{equation}
\Phi\circ\cmss R=\Phi,\qquad \cmss R\in \SO(n).
\end{equation}
(2) For any~$x,y$ we have $K_{xy}\ge0$ and $K_{xy}=K_{0,y-x}$. Moreover, there is an~$s>d$ such that
\begin{equation}
\label{K-asympt}
0<\liminf_{|x|\to\infty}|x|^s K_{0x}\le\limsup_{|x|\to\infty}|x|^s K_{0x}<\infty
\end{equation}
Here (and henceforth) $|x|$ denotes the Euclidean norm of~$x$.
\end{assumption}


Let~$r$ be as in Assumption~\ref{ass}(1). The Hamiltonian~$\scrH_N$ in~$\Lambda_N$ is then defined by
\begin{equation}
\label{E:Ham}
\scrH_N(\sigma):=\sum_{x\in\Lambda_{N+r}}
\Phi\circ\tau_x(\sigma)
+\,\lambda\!\!\!\sum_{\begin{subarray}{c}
x,y\colon x\ne y\\\{x,y\}\cap\Lambda_N\ne\emptyset
\end{subarray}}
\!\!K_{xy}\,\sigma_x\cdot\sigma_y.
\end{equation}
The conditions (1) and (2) ensure that the interaction is well-defined, shift-invariant --- and so it will make sense to talk about translation invariant and ergodic Gibbs measures --- and also invariant under simultaneous (rigid) rotations of all spins, i.e., $\scrH_N(\cmss R\sigma)=\scrH_N(\sigma)$. The model \eqref{1.1} is clearly a special case of \eqref{E:Ham}.

We will need to invoke the formalism of infinite-volume Gibbs measures for which we refer the reader to the standard treatments by Georgii~\cite{Georgii} or Simon~\cite{Simon}. We will only mention the features that are relevant for our problem.
Let~$\nu$ denote the uniform probability (Haar) measure on~$\BbbS^{n-1}$. The above Hamiltonian defines a finite-volume Gibbs specification~$\gamma_N$ with boundary condition~$\bar\sigma\in\Omega$ via
\begin{equation}
\gamma_N(\textd\sigma|\bar\sigma):=\frac1{Z_N(\bar\sigma)}\,\texte^{-\beta\scrH_N(\sigma)}\prod_{x\in\Lambda_N}\nu(\textd\sigma_x)\prod_{z\not\in\Lambda_N}\delta_{\bar\sigma_z}(\textd\sigma_z).
\end{equation}
Here $\beta\ge0$ denotes the inverse temperature, $Z_N(\bar\sigma)$ is the partition function and~$\delta_{\bar\sigma_z}$ is the Dirac point mass at $\bar\sigma_z$. We say that a probability measure~$\mu$ over~$\Omega$ is a \emph{Gibbs measure}, if for all events~$A$ and all~$N\ge1$,
\begin{equation}
\label{E:Gibbs}
E_\mu(\gamma_N(A|\cdot))=\mu(A).
\end{equation}
Here~$E_\mu$ denotes expectation with respect to~$\mu$.
We say that a measure~$\mu$ is \emph{translation invariant} if $\mu\circ\tau_x=\mu$ for all~$x\in\Z^d$. The measure is \emph{ergodic} if~$\mu(A)=0$ or~$1$ for all events~$A$ such that~$\tau_x(A)=A$ for all~$x\in\Z^d$.

In order to define the notion of the spontaneous magnetization, pick a unit vector $\hate\in\R^n$ and consider the function~$f\colon\R\to\R$ defined by
\begin{equation}
\label{E:f}
f(h):=\lim_{N\to\infty}\sup_{\bar\sigma}\frac1{|\Lambda_N|}\log\int\prod_{x\in\Lambda_N}\nu(\textd\sigma_x)\exp\Bigl\{-\beta\scrH_N(\sigma)+h\sum_{x\in\Lambda_N}\hate\cdot\sigma_x\Bigr\},
\end{equation}
where $\sigma_x$ is (implicitly) fixed to~$\bar\sigma_x$ for any~$x\not\in\Lambda_N$ on which~$\scrH_N(\sigma)$ depends. The limit exists by subadditivity arguments and is convex as a function of~$h$. In addition, by the invariance of~$\scrH_N$ and the measure~$\nu$ with respect to rotations, $f$ is independent of the choice of~$\hate \in \mathbb S^{n-1}$. The convexity of~$f$ ensures the existence of the right derivative
\begin{equation}
\label{E:2.7}
m_\star:=\frac{\textd}{\textd h^+}f(h)\Bigl|_{h=0},
\end{equation}
which is non-negative by symmetry $\hate\leftrightarrow-\hate$. We will call $m_\star$ the \emph{spontaneous magnetization}.

As is well known (see, e.g., Theorem~2.3(3) of \cite{Biskup}), for each unit vector~$\hate$ there is a translation-invariant (and, in fact, ergodic) Gibbs measure~$\mu$ such that $E_\mu(\sigma_x)=m_\star\hate$. Note that, in light of our remarks from Section~\ref{sec1}, we are \emph{not} assuming that this~$\mu$ is \emph{extremal}, which would mean that $\mu(A)=0$ or~$1$ for any event~$A$ that does not depend on the state of any finite number of~$\sigma_x$'s. (Note that, by the Backward Martingale Limit Theorem, extremality implies clustering \eqref{E:cluster}).

\smallskip
Our main result is now the following:

\begin{theorem}
\label{thm1}
Suppose~$d\ge1$, $n\ge2$ and~$s\in(d,d+2]$ and consider a model satisfying Assumptions~\ref{ass}. Then for any $\lambda>0$ and any inverse temperature~$\beta\ge0$,
\begin{equation}
\label{E:mu-mstar}
E_\mu\sigma_x=0,\qquad x\in\Z^d,
\end{equation}
holds for every translation-invariant Gibbs measure~$\mu$.
In particular, the spontaneous magnetization vanishes, i.e.,~$m_\star=0$, and~$h\mapsto f(h)$ is differentiable at~$h=0$ with $f'(0)=0$.
\end{theorem}

This statement is restricted to translation-invariant Gibbs measures. But a version of this conclusion is possible for all Gibbs measures:

\begin{corollary}
\label{cor1}
Under the conditions of Theorem~\ref{thm1}, if~$\mu$ is any Gibbs measure, then
\begin{equation}
\lim_{N\to\infty}\frac1{|\Lambda_N|}\sum_{x\in\Lambda_N}\sigma_x=0,\qquad \mu\text{\rm-a.s.},
\end{equation}
i.e., block-averages of the spins tend to zero in almost every sample from~$\mu$.
\end{corollary}

Note that these results do not preclude other types of long-range order (e.g., stripe states or an orientational order). A few additional remarks are in order:

\begin{remarks}
\label{R}
(1) The proof we construct below would work even if we assumed that $\sigma\mapsto\Phi(\sigma)$ --- as a function on $(\BbbS^{n-1})^{\Lambda_r}$ --- has only a Lipschitz-continuous derivative. However, we suspect that the theorem holds even when~$\Phi$ is just continuous. This would be analogous to a strong form of the Mermin-Wagner theorem proved by Ioffe, Shlosman and Velenik~\cite{ISV}.

(2) The positivity requirement for $K_{xy}$ comes at no loss as finite-range deviations from this can be absorbed into the short-range part of the interaction. The requirement that $K_{xy}$ be the order of $|x-y|^{-s}$ can, for~$s<d+2$, be replaced by $K_{0x}=|x|^{-s+o(1)}$ (for $|x|\to\infty$). However, the explicit form of the $o(1)$ term becomes relevant in the boundary case $s=d+2$. In this context it might be of interest to see whether a sharp (e.g., summability) condition exists implying absence of magnetic order. Even the translation invariance of $K_{xy}$ is not essential for the proofs as long as the tail estimate can be made uniform in the position.

(3) Our statement and proof rule out ferromagnetic order, but more subtle, e.g., antiferromagnetic ordering is not excluded at all. It is, in fact, quite possible that antiferromagnetic nearest-neighbor coupling in $d\ge3$, or more subtle, order-by-disorder induced antialignment in $d=2$ (e.g., as in~\cite{BCK2}) will persist for all exponents~$s>d$ as long as the overall strength of the long-range interaction is small. On the other hand, just as interesting would be to show that, for all $s>d+2$, the ferromagnetic order exists for the model \eqref{1.1} once $0<\lambda\ll J$ and $\beta$ sufficiently large. We note that the principal proof technique, reflection positivity, seems to fail in this case.

(4) Our method of proof is related to uniqueness arguments by Bricmont, Lebowitz and Pfister~\cite{BLP} as well as the deformation arguments underpinning the Mermin-Wagner theorem (see, e.g., Simon~\cite{Simon} or, more specifically, Bonato, Perez and Klein \cite[Theorem~1]{Bonato-Perez-Klein}). Combining our results with these, we can strengthen the conclusions as follows:
\begin{itemize}
\item
For $d=1$, $s>2$ the Gibbs measure is unique.
\item
For $d=1$ \& $s=2$ or $d=2$ \& $s=4$, the Gibbs measure is globally $\justO(n)$-invariant.
\item
For $d\ge1$ AND $d<s\le d+2$, the magnetization vanishes.
\end{itemize}
A few questions remain: Is it possible that the Gibbs measure is actually unique for $1<s<2$ ($d=1$) as well?
Or is it possible that, in dimensions $d=1,2$, the Gibbs measures are $\justO(n)$-invariant for all $d<s<d+2$?

(5) Turning to the question of stripe or modulated order in these systems, a natural first case to address would be an appropriate Kac limit of the type discussed in~\cite{Gates-Penrose,Pisani-Smith-Thompson,Pisani-Thompson}. However, compared to the models treated in these references, in the present model the short and long-range parts of the interaction appears to be characterized by different scaling dimensions and so it is not immediately how to properly Kac-ify the short-range part of the Hamiltonian.

(6) Finally, it would be of much interest to see whether any of the present methods can be extended to the case of non-scalar interaction --- e.g., the dipole-dipole model mentioned in the introduction. The dependence of the \emph{sign} of the interaction on the relative orientation of the spins to their mutual positions is one of the key issues to overcome here.
\end{remarks}

\section{Main steps of the proof}
\label{sec3}\noindent
Suppose $n\ge2$ and fix a potential~$\Phi$, the coupling constants~$K_{xy}$ and exponent $s\in(d,d+2]$ so that Assumptions~\ref{ass} hold. Pick constants $\lambda>0$ and~$\beta>0$. We will assume that~$m_\star>0$ and derive a contradiction. Let~$\hate_i$ denote the~$i$-th coordinate vector in~$\R^n$ and let~$\mu$ denote  a translation-invariant Gibbs measure for which we have
\begin{equation}
\label{E:mu}
E_\mu(\sigma_x)=m_\star\hate_1,\qquad x\in\Z^d.
\end{equation}
As already mentioned, this measure exists by Theorem~2.3(3) of \cite{Biskup}.

Now we pick two length scales $L$ and~$a$ taking values in~$\D:=\{2^k\colon k\in\N\}$ with~$L>a$, and consider a deformation of the spin configuration inside~$\Lambda_L$ that reverts the orientation of the first two components of the spin everywhere inside~$\Lambda_{L-a}$. Explicitly, let $\cmss R\in \SO(n)$ be the rotation such that $\cmss R\hate_i=-\hate_i$ for~$i=1,2$ while $\cmss R\hate_i=\hate_i$ for~$i>2$. We can view~$\cmss R$ as the endpoint of a continuous trajectory of maps
\begin{equation}
\label{E:Rtheta}
\cmss R^\theta:=\left(\begin{matrix}
\cos\theta&\sin\theta&0&\cdots&0\\
-\sin\theta&\cos\theta&0&\cdots&0\\
0&0&1&\cdots&0\\
\vdots&\ddots&\ddots&\ddots&\vdots\\
0&0&0&\cdots&1\\
\end{matrix}\right)
\end{equation}
as $\theta$ varies either from $0$ to~$\pi$ or from $0$ to~$-\pi$.
Next we define the ``deformation angles''
\begin{equation}
\theta_x:=
\begin{cases}
\pi a^{-1}\dist(x,\Lambda_L^\cc),\qquad&\text{if }x\not\in\Lambda_{L-a},
\\
\pi,\qquad&\text{if }x\in\Lambda_{L-a},
\end{cases}
\end{equation}
where~$\dist(x,y)$ is the $\ell^\infty$-distance on~$\Z^d$. These permit us to define the global \emph{inhomogeneous} rotations~$\cmss R^\pm$ on the configuration space by
\begin{equation}
\label{E:Rpm}
(\cmss R^\pm\sigma)_x:=\cmss R^{\pm\theta_x}\sigma_x,\qquad x\in\Z^d.
\end{equation}
Notice that~$(\cmss R^\pm\sigma)_x=-\sigma_x$ for~$x\in\Lambda_{L-a}$ while $(\cmss R^\pm\sigma)_x=\sigma_x$ for~$x\in\Lambda_L^\cc$.

Our next step will be to use these rotations to express quantitatively the assumption of differentiability of the map~$\Phi$. Abusing the notation slightly, let $\cmss R_x^\varphi$ denote the inhomogeneous rotation of~$\sigma$ such that $(\cmss R_x^\varphi\sigma)_z=\sigma_z$ when~$z\ne x$ and $(\cmss R_x^\varphi\sigma)_x=\cmss R^\varphi\sigma_x$. The map~$\varphi \mapsto \Phi(\cmss R_x^\varphi\sigma)$ is differentiable and the corresponding derivative is
\begin{equation}
D_x\Phi(\sigma):=\frac{\textd}{\textd \varphi}\Phi(\cmss R_x^\varphi\sigma)\Bigr|_{\varphi=0}.
\end{equation}
Similarly, $D_xD_y\Phi(\sigma):=D_x(D_y\Phi)(\sigma)$. We also write
\begin{equation}
\label{E:norm}
\Vert\Phi''\Vert:=\sup_{\begin{subarray}{c}
w\in\ell^2(\Z^d)\\\Vert w\Vert_2=1
\end{subarray}}
\sup_\sigma\,\biggl|\,\sum_{z,z'}w_zw_{z'}D_zD_{z'}\Phi(\sigma)\biggr|
\end{equation}
to denote a natural norm of the second derivative of~$\Phi$. We remark that, while the derivatives are defined using a specific one-parameter subgroup~$\theta\mapsto\cmss R^\theta$ of~$\SO(n)$, the rotation invariance of~$\Phi$ makes the specific choice of the subgroup immaterial.

Suppose now that~$N>L+r$ and $L>a$. The entire argument is centered around the probability distribution of the \emph{energy defect}
\begin{equation}
\Delta_{L,a}(\sigma):=2\scrH_N(\sigma)-\scrH_N(\cmss R^+\sigma)-\scrH_N(\cmss R^-\sigma).
\end{equation}
Notice that this quantity is independent of~$N$ as long as $N>L+r$. The reasons for consideration of both~$\cmss R^+$ and~$\cmss R^-$  --- inspired by some proofs of the Mermin-Wagner theorem (e.g.,~Fr\"ohlich and Pfister~\cite{Frohlich-Pfister}) and employed also by van Enter~\cite{vanEnterPRB} --- will become very apparent from the proof of a uniform bound on~$\Delta_{L,a}$:

\begin{lemma}
\label{L:abs-bound}
For all $L>a$ and all~$\sigma\in\Omega$,
\begin{equation}
\bigl|\Delta_{L,a}(\sigma)\bigr|\le U_{L,a},
\end{equation}
where
\begin{equation}
\label{E:U-def}
U_{L,a}:=\Vert\Phi''\Vert\sum_{x\in\Z^d}\sum_{y\in\Lambda_r}
(\theta_{x+y}-\theta_x)^2
+\!|\lambda|\sum_{x,y\colon x\ne y}
\!\!K_{xy}\,(\theta_x-\theta_y)^2.
\end{equation}
\end{lemma}

\begin{proofsect}{Proof}
We will first deal with the long-range part of the interaction. Let~$\varphi_x$ denote the polar angle for the projection of~$\sigma_x$ onto the subspace of~$\R^n$ spanned by~$\hate_1$ and~$\hate_2$, and let~$s_x$ denote the projection of~$\sigma_x$ onto the subspace of~$\R^n$ spanned by~$\hate_i$, $i=3,\dots,n$. Then
\begin{equation}
\sigma_x\cdot\sigma_y=s_x\cdot s_y+\sqrt{1-s_x^2}\,\sqrt{1-s_y^2}\,\cos(\varphi_x-\varphi_y).
\end{equation}
Since the rotation of the spins occurs only in the $\hate_1,\hate_2$-plane, $s_x$ is not changed when~$\cmss R^\theta$ is applied to~$\sigma_x$. Therefore,
\begin{multline}
\qquad
\bigl|2\sigma_x\cdot\sigma_y-(\cmss R^+\sigma)_x\cdot(\cmss R^+\sigma)_y-(\cmss R^-\sigma)_x\cdot(\cmss R^-\sigma)_y\bigr|
\\
\le\bigl|2\cos(\varphi_x-\varphi_y)-\cos(\varphi_x-\varphi_y+\theta_x-\theta_y)
\\
-\cos(\varphi_x-\varphi_y-\theta_x+\theta_y)\bigr|.
\qquad
\end{multline}
It is now easy to check that the right hand side is no larger than~$(\theta_x-\theta_y)^2$. Using this for all long-range terms in~$\Delta_{L,a}$, we get the second term in \eqref{E:U-def}.

In order to control the short-range contribution to~$\Delta_{L,a}$, note that for each $x$, we can use invariance of $\Phi$ under $\SO(n)$ to write the corresponding term in the interaction as
\begin{equation}
2\Phi\circ\tau_x(\sigma)-\Phi\circ \tau_x \bigl(\cmss R^{-\theta_{x}} \cmss R^+ \sigma \bigr)-\Phi\circ\tau_x\bigl(\cmss R^{\theta_{x}} \cmss R^-\sigma \bigr).
\end{equation}
Abbreviate $\vartheta_z:=\theta_z-\theta_x$ and, for $t\in[-1,1]$, let $\cmss S^t$ denote the composition of the maps~$\cmss R_z^{t\vartheta_z}$ for all~$z$. Then $\cmss R^{-\theta_{x}} \cmss R^+=\cmss S^1$ and $\cmss R^{\theta_{x}} \cmss R^-=\cmss S^{-1}$ and, for $\Psi:=\Phi\circ\tau_x$,
\begin{equation}
\label{E:Psi}
2\Psi(\sigma)-\Psi(\cmss S^1\sigma)-\Psi(\cmss S^{-1}\sigma)
=-\int_{0}^1\textd t\int_{-t}^t\textd u\,\sum_{z,z'}\vartheta_z\vartheta_{z'}D_zD_{z'}\Psi(\cmss S^{u}\sigma).
\end{equation}
The integrand is now bounded via
\begin{equation}
\biggl|\,\sum_{z,z'}\vartheta_z\vartheta_{z'}D_zD_{z'}\Psi(\cmss S^{u}\sigma)\biggr|
\le\Vert\Phi''\Vert\sum_{z\in\Lambda_r}\vartheta_{x+z}^2,
\end{equation}
where we used that $D_zD_{z'}\Psi(\cmss S^{u}\sigma)=0$ unless $z-x,z'-x\in\Lambda_r$.
The integral over~$u$ and~$t$ then gives a factor of one; the claim then follows by summing the result over~$x$.
\end{proofsect}

Our next observation will be concerned with the leading-order growth of~$U_{L,a}$.

\begin{proposition}
\label{P:UI-bd}
Assume~$\lambda>0$. For each $d\ge1$, $s\in(d,d+2]$ and each value of the ratio $\Vert\Phi''\Vert/\lambda$ there is a constant~$c\in(0,1)$ such that if~$c^{-1}\le a\le cL$, then
\begin{equation}
\label{E:3.10}
c\lambda \II_{L,a}\le U_{L,a}\le c^{-1}\lambda \II_{L,a},
\end{equation}
where
\begin{equation}
\label{E:I-def}
\II_{L,a}:=
L^{d-1} \times
\begin{cases}
L^{d+1-s},\qquad&\text{if }\,d<s<d+1,
\\
\log(\ffrac La),\qquad&\text{if }\,s=d+1,
\\
a^{d+1-s},\qquad&\text{if }\,d+1<s<d+2,
\\
a^{-1}\log a,\qquad&\text{if }\,s=d+2.
\end{cases}
\end{equation}
\end{proposition}

The proof of these bounds is relatively straightforward but, in order to stay focused on the main line of argument, we defer it to Section~\ref{sec4}. The quantity  $\II_{L,a}$ will play the role of a benchmark scale for all arguments that are to follow.
Our next step is the connection between the above energy defect and positive magnetization:

\begin{proposition}
\label{P:Delta-mstar}
Suppose~$m_\star>0$ and let~$\mu$ be an ergodic Gibbs measure satisfying \eqref{E:mu}. For each $\kappa\in(0,m_\star^2)$ there is~$c'\in(0,1)$ such that if $a,L\in\D$ obey $1/c'<a\le c' L$, then
\begin{equation}
E_\mu\bigl(\Delta_{L,a}(\sigma)\bigr)
\ge c'(m_\star^2-\kappa) \,\II_{L,a}.
\end{equation}
\end{proposition}

Again, to keep the main argument free of lengthy technical interruptions, we postpone the proof to Section~\ref{sec5}. This estimate enters the main argument via:

\begin{lemma}
\label{L:E-P}
Fix~$\lambda>0$ and let~$c\in(0,1)$ be the constant from Proposition~\ref{P:UI-bd}. Suppose that~$c^{-1}\le a\le cL$. Then for each~$\zeta\in[0,c^{-1}\lambda)$,
\begin{equation}
\mu(\Delta_{L,a}\ge\zeta\II_{L,a})\ge\frac{E_\mu(\Delta_{L,a})-\zeta\II_{L,a}}{(c^{-1}\lambda-\zeta) \II_{L,a}}.
\end{equation}
\end{lemma}

\begin{proofsect}{Proof}
The absolute bound from Lemma~\ref{L:abs-bound} tells us
\begin{equation}
E_\mu(\Delta_{L,a})\le\mu(\Delta_{L,a}\ge\zeta\II_{L,a})\bigl[U_{L,a}-\zeta\II_{L,a}\bigr]+\zeta\II_{L,a}.
\end{equation}
Using \eqref{E:3.10} and~$\zeta<c^{-1}\lambda$, the claim now easily follows.
\end{proofsect}

The last essential ingredient we will need is the following fact:

\begin{lemma}
\label{L:exp-bd}
For each~$L>a$, any event~$A$ depending only on~$\{\sigma_x\colon x\in\Lambda_{L-a}\}$, any Gibbs measure~$\mu$ and any~$t\in\R$ we have
\begin{equation}
\mu\bigl(A\cap\{\Delta_{L,a}\ge t\}\bigr)\le\texte^{-\frac12\beta t}\mu\bigl(\cmss R(A)\bigr).
\end{equation}
\end{lemma}

\begin{proofsect}{Proof}
Let~$N>L+r$ and abbreviate $A_t:=A\cap\{\Delta_{L,a}\ge t\}$. Then for any~$\sigma\in A_t$,
\begin{equation}
\texte^{-\beta\scrH_N(\sigma)}\le\texte^{-\frac12\beta t}\,\texte^{-\frac12\beta\scrH_N(\cmss R^+\sigma)-\frac12\beta\scrH_N(\cmss R^-\sigma)}.
\end{equation}
It follows that
\begin{equation}
\gamma_N(A_t|\bar\sigma)
\le\frac{\texte^{-\frac12\beta t}}{Z_N(\bar\sigma)}
\int_{A_t}
\,\texte^{-\frac12\beta\scrH_N(\cmss R^+\sigma)-\frac12\beta\scrH_N(\cmss R^-\sigma)}
\prod_{x\in\Lambda_N}\nu(\textd\sigma_x),
\end{equation}
where we think of all~$\sigma_x$ with~$x\not\in\Lambda_N$ as fixed to~$\bar\sigma_x$. Using the Cauchy-Schwarz inequality and $A_t\subset A$, the last integral is bounded by the product
\begin{equation}
\biggl(\int_{A}
\,\texte^{-\beta\scrH_N(\cmss R^+\sigma)}\prod_{x\in\Lambda_N}\nu(\textd\sigma_x)
\biggr)^{\ffrac12}\biggl(\int_{A}
\,\texte^{-\beta\scrH_N(\cmss R^-\sigma)}\prod_{x\in\Lambda_N}\nu(\textd\sigma_x)
\biggr)^{\ffrac12}.
\end{equation}
But~$\cmss R^\pm$ alter only the spins inside~$\Lambda_L$ and since the product measure is~$\cmss R^\pm$-invariant, and $\cmss R^\pm(A)=\cmss R(A)$, both integrals are equal to~$Z_N(\bar\sigma)\gamma_L(\cmss R(A)|\bar\sigma)$. Therefore,
\begin{equation}
\gamma_N(A_t|\bar\sigma)\le\texte^{-\frac12\beta t}\gamma_L(\cmss R(A)|\bar\sigma),\qquad\bar\sigma\in\Omega.
\end{equation}
The claim is now proved by taking expectation with respect to~$\mu$.
\end{proofsect}

Now we are ready to begin the actual proof of our main result. We need to observe:

\begin{lemma}
\label{L:I-diverge}
Suppose either $d\ge2$ {\rm AND} $s\in(d,d+2]$ or $s=1$ {\rm AND} $1<s\le2$. Then there is a way to take $L,a\to\infty$ so that~$L/a\to\infty$ and $\II_{L,a}\to\infty$.
\end{lemma}

\begin{proofsect}{Proof}
This is directly verified from the formula \eqref{E:I-def}.
\end{proofsect}

\begin{proofsect}{Proof of Theorem~\ref{thm1}}
Suppose~$d\ge1$ and~$s\in(d,d+2]$. If $d=1$ and $s>2$, then the interaction \eqref{E:Ham} satisfies the conditions of Corollary~1 in Bricmont, Lebowitz and Pfister~\cite{BLP} which implies that the Gibbs state is unique and so there is nothing to prove. We may thus assume that $1<s\le2$ in $d=1$ and so the conclusion of Lemma~\ref{L:I-diverge} is available in all cases of concern. Let~$\lambda>0$ and suppose~$m_\star>0$. Note that this also permits us to assume~$\beta>0$.  Let~$\mu$ be a translation-invariant Gibbs measure obeying \eqref{E:mu-mstar} and let~$c$ be as in Proposition~\ref{P:UI-bd}. Pick~$\kappa\in(0,m_\star^2)$ and let~$c'\in(0,1)$ be as in Proposition~\ref{P:Delta-mstar}. Set~$c'':=\min\{c,c'\}$ and suppose~$L,a\in\D$ are such that $1/c''\le a\le c''L$ for the rest of the argument.

Fix $\zeta$ such that $0<\zeta<c'(m_\star^2-\kappa)$ and~$\zeta<c^{-1}\lambda$. Proposition~\ref{P:Delta-mstar} and Lemmas~\ref{L:E-P}-\ref{L:exp-bd} yield
\begin{equation}
\texte^{-\frac12\beta\zeta\II_{L,a}}\ge
\mu(\Delta_{L,a}\ge\zeta\II_{L,a})\ge\frac{c'(m_\star^2-\kappa)-\zeta}{c^{-1}\lambda-\zeta}>0.
\end{equation}
But this leads to a contradiction because Lemma~\ref{L:I-diverge} permits us to take~$L,a\to\infty$ --- subject to the aforementioned restrictions --- so that  $\II_{L,a}\to\infty$. Hence~$m_\star=0$ as claimed.

The conclusion~$m_\star=0$ implies that the derivative in \eqref{E:2.7} vanishes and since~$h\mapsto f(h)$ is even,~$f$ is differentiable at $h=0$ with, inevitably, $f'(0)=0$. That this implies \eqref{E:mu-mstar} is the consequence of standard thermodynamic arguments (see, e.g.,~\cite[Theorem~2.5(2)]{Biskup}); we spell these out for convenience in the next proof.
\end{proofsect}

\begin{proofsect}{Proof of Corollary~\ref{cor1} and \eqref{E:mu-mstar}}
Fix~$\delta>0$ and pick a unit vector~$\hate\in\R^n$. By the exponential Chebyshev inequality, for any~$h>0$,
\begin{equation}
\label{E:sigma-dev}
\mu\Bigl(\,\sum_{x\in\Lambda_N}\sigma_x\cdot\hate>\delta|\Lambda_N|\Bigr)
\le\texte^{-h\delta|\Lambda_N|}E_\mu\biggl(\,\exp\Bigl\{h\sum_{x\in\Lambda_N}\sigma_x\cdot\hate\Bigr\}\biggr).
\end{equation}
Invoking \eqref{E:Gibbs}, the definition \eqref{E:f} then yields
\begin{equation}
\limsup_{N\to\infty}\frac1{|\Lambda_N|}\log\mu\Bigl(\,\sum_{x\in\Lambda_N}\sigma_x\cdot\hate>\delta|\Lambda_N|\Bigr)
\le f(h)-h\delta.
\end{equation}
But~$f(h)=o(h)$ as~$h\downarrow0$ by the fact that~$m_\star=0$ and so $f(h)-h\delta<0$ once~$h$ is sufficiently small. The probability in \eqref{E:sigma-dev} thus decays exponentially in~$|\Lambda_N|$ and so, by Borel-Cantelli, the corresponding event occurs only for finitely many~$N$, $\mu$-a.s. As this holds for all~$\delta>0$ and all~$\hate$, the block-average of spins is zero in any Gibbs measure. For translation invariant~$\mu$, Fatou's lemma (or the Ergodic Theorem) then imply \eqref{E:mu-mstar}.
\end{proofsect}

\begin{remark}
\label{R-MW}
The above proof uses Lemma~\ref{L:exp-bd} only for $A:=\Omega$ but we introduced the more general statement as it can be used to establish a Mermin-Wagner type result. Indeed, suppose that we can take $L\to\infty$ while adjusting~$a$ (with $1\ll a\ll L$) so that $\II_{L,a}$ stays bounded from above, say $\II_{L,a}\le t$. By Lemma~\ref{L:abs-bound} and Proposition~\ref{P:UI-bd} we have $\Delta_{L,a}\ge-c^{-1}\lambda t$ and so
\begin{equation}
\mu\bigl(\cmss R(A)\bigr)\ge\texte^{-\frac12\beta c^{-1}\lambda t}\mu(A)
\end{equation}
for all Gibbs states and, via standard extension arguments, all events~$A$. Applying this to extremal Gibbs states and extremal events (where all probabilities are either zero or one) we immediately conclude that $\mu\circ\cmss R=\mu$, i.e., $\mu$ is invariant under the rotation by $180^\circ$ in the subspace of the spin space spanned by $\hate_1,\hate_2$. But our choice of the basis in the spin space was arbitrary and so the Gibbs state $\mu$ is invariant under all such rotations in~$\justO(n)$. These rotations generate (via Euler angles) the whole group and so $\mu$ is globally $\justO(n)$-invariant.

It remains to check the conditions under which $\II_{L,a}$ remains bounded in the limit $L\to\infty$ when $a=a_L$ satisfies $1\ll a_L\ll L$. A glance at \eqref{E:I-def} reveals that this can be done for $s\in[2,3]$ in $d=1$ and $s=4$ in $d=2$. For $d=1$ and $s>2$ we already know the Gibbs measure is unique, and so the conclusion is non-trivial for $d=1$ \& $s=2$ and $d=2$ \& $s=4$. But this is exactly covered by the general theory on the Mermin-Wagner theorem (e.g., Theorem~1 of Bonato, Perez and Klein \cite{Bonato-Perez-Klein}). 
\end{remark}

\section{Estimates on interaction strength}
\label{sec4}\noindent
In this section we will prove Proposition~\ref{P:UI-bd} by analyzing the various contributions to the quantity~$U_{L,a}$, which serves as the uniform upper bound on the energy defect. To keep our notations succinct, we will write this quantity as
\begin{equation}
U_{L,a}=\Vert\Phi''\Vert Q^++|\lambda|Q^-,
\end{equation}
where
\begin{equation}
Q^+:=\sum_{x\in\Z^d}\sum_{y\in\Lambda_r}
|\theta_{x+y}-\theta_x|^2
\end{equation}
and
\begin{equation}
\label{Qminus}
Q^-:=\sum_{x,y\colon x\ne y}K_{xy}|\theta_x-\theta_y|^2.
\end{equation}
Then we have the following estimates:

\begin{lemma}
\label{L:Aplus}
There is~$c_1\in(0,1)$ such that for all $L,a$ with $c_1^{-1}\le a\le c_1L$,
\begin{equation}
Q^+\le c_1L^{d-1}a^{-1}.
\end{equation}
\end{lemma}

\begin{lemma}
\label{L:Aminus}
Let~$s\in(d,d+2]$. There is $c_2\in(0,1)$ such that for all $L,a$ with $c_2^{-1}\le a\le c_2 L$,
\begin{equation}
\label{E:I-bounds}
c_2\II_{L,a}\le Q^-\le c_2^{-1}\II_{L,a},
\end{equation}
where~$\II_{L,a}$ as in \eqref{E:I-def}.
\end{lemma}

Let us first see how this yields the desired asymptotic for $U_{L,a}$:

\begin{proofsect}{Proof of Proposition~\ref{P:UI-bd}}
Notice that for all~$s\le d+2$ the ratio $L^{d-1}a^{-1}/\II_{L,a}$ tends to zero in the limit when~$L,a\to\infty$ with~$a/L\to0$ and so we can easily arrange that $Q^+/Q^-$ is arbitrarily small by making~$a$ and $L/a$ large enough. The claim follows.
\end{proofsect}

It remains to prove the two lemmas above. The first one is easy:

\begin{proofsect}{Proof of Lemma~\ref{L:Aplus}}
We have $|\theta_{x+y}-\theta_x|\le\pi ra^{-1}$ for $x\in\Lambda_{L+r}\setminus\Lambda_{L-r-a}$ and $y\in\Lambda_r$, while the difference is zero (or is irrelevant) in other cases. Consequently, the sum is at most of order $r^2(2r+a)(L+r)^{d-1}a^{-2}$. As~$r$ is fixed, this readily yields the claim.
\end{proofsect}

For the proof of Lemma~\ref{L:Aminus}, we will need to introduce some additional notation. First, the contributions to $Q^-$ can be divided into four different categories depending on the containments of $x$ and $y$ in \eqref{Qminus} in $\Lambda_L$ and~$\Lambda_{L-a}$. We introduce four sets of relevant pairs:
\begin{equation}
\begin{aligned}
\PP_1&:=\bigl\{(x,y)\colon x\in\Lambda_{L-a},\,y\in\Lambda_{L}\setminus\Lambda_{L-a}\bigr\},
\\
\PP_2&:=\bigl\{(x,y)\colon x\in\Lambda_{L-a},\,y\in\Lambda_{L}^\cc\bigr\},
\\
\PP_3&:=\bigl\{(x,y)\colon x,y\in\Lambda_{L}\setminus\Lambda_{L-a},\,|x|_\infty<|y|_\infty\bigr\},
\\
\PP_4&:=\bigl\{(x,y)\colon x\in\Lambda_{L}\setminus\Lambda_{L-a},\,y\in\Lambda_{L}^\cc\bigr\}
\end{aligned}
\end{equation}
and use these to define
\begin{equation}
\label{E:4.7}
Q_i^-:=\sum_{(x,y)\in\PP_i}(K_{xy}+K_{yx})|\theta_y-\theta_x|^2,\qquad i=1,\dots,4.
\end{equation}
Since the deformation angles are constant on $\Lambda_{L-a}$ and on $\Lambda_L^\cc$, and $\theta_x=\theta_y$ when $|x|_\infty=|y|_\infty$, we easily convince ourselves that
\begin{equation}
\label{E:4.8}
Q^-=Q_1^-+Q_2^-+Q_3^-+Q_4^-.
\end{equation}
It thus suffices to provide the relevant estimates on $Q_i^-$ alone. In order to do so, we will introduce yet simpler quantities $q_1,\dots,q_4$ that capture the essential contributions to $Q_1^-,\dots,Q_4^-$ modulo a ``surface'' term of order~$L^{d-1}$.

Explicitly, our bound on $Q_1^-$ will boil down to estimating the quantity
\begin{equation}
\label{Eq1}
q_1:=\sum_{u=0}^L\sum_{t=1}^a\sum_{z\in \Z^{d-1}}\frac{(\ffrac ta)^2}{[(u+t)^2+|z|^2]^{s/2}}.
\end{equation}
For $Q_2^-$, which will turn out to be one of two dominant terms, we introduce the notation
\begin{equation}
R_N:=\bigl\{z=(z_1,\dots,z_{d-1})\in\Z^{d-1}\setminus\{0\}\colon |z_i|\le N/2\bigr\}.
\end{equation}
Our bounds on $Q_2^-$ will then be expressed in terms of
\begin{equation}
q_{2,N}:=\sum_{u=0}^L\sum_{t> a}\sum_{z\in R_N}\frac{1}{[(u+t)^2+|z|^2]^{s/2}}
\end{equation}
for $N:=L$ and $N:=\infty$. As to $Q_3^-$, we will similarly need
\begin{equation}
q_{3,N}:=\sum_{u=0}^a\sum_{t=u+1}^a\sum_{z\in R_N}\frac{\bigl(\tfrac{t-u}a\bigr)^2}{[(t-u)^2+|z|^2]^{s/2}}.
\end{equation}
Finally, a control of $Q_4^-$ will require bounding
\begin{equation}
\label{Eq4}
q_4:=\sum_{u=0}^a\sum_{t\ge0}\sum_{z\in \Z^d}\frac{(\ffrac ua)^2}{[(u+t)^2+|z|^2]^{s/2}}.
\end{equation}
The connection of these quantities to $Q^-$ is provided by:

\begin{lemma}
\label{lemma4.3}
Let $s\in(d,d+2]$ and assume $a\le L/2$. Then there are constants $c=c(s,d)\in(0,1)$ and $c'\in(0,\infty)$, with~$c'$ depending only on $(K_{xy})$, such that
\begin{equation}
Q^-\le c^{-1}L^{d-1}(q_1+q_{2,\infty}+q_{3,\infty}+q_4)+c'L^{d-1}a^{-1}
\end{equation}
and
\begin{equation}
Q^-\ge cL^{d-1}(q_{2,L}+q_{3,L})-c'L^{d-1}a^{-1}.
\end{equation}
\end{lemma}

Before we set out to give a proof, we need to make a  geometric observation. For vertices $x=(x_1,\dots,x_d)\in\Z^d\setminus\{0\}$ and $y=(y_1,\dots,y_d)\in\Z^d$ we define a vertex $T_x(y)$ as follows: Let $i$ denote the smallest index such that $|x_i|=\max_k|x_k|$ and let $j$ denote the smallest index such that $|y_j|=\max_k|y_k|$. Then $T_x(y):=(\tilde y_1,\dots,\tilde y_d)$ for $\tilde y_k:=y_k$ when $k\ne i,j$, 
\begin{equation}
\tilde y_i:=\sign(x_i)|y_i|\quad\text{when}\quad i=j
\end{equation}
and
\begin{equation}
(\tilde y_i,\tilde y_j):=\bigl(\sign(x_i)|y_j|,\sign(x_iy_j)|y_i|\bigr)
\quad\text{when}\quad i\ne j.
\end{equation}
Notice that $T_x$ maps~$\Z^d$ into the ``wedge'' $\{(z_1,\dots,z_d)\in\Z^d\colon |z_i|=\max_k|z_k|,\,\sign(z_i)=\sign(x_i)\}$ and each vertex there has at most $2d$ preimages. The principal fact about this map is:

\begin{lemma}
\label{lemma4.4}
For any $x,y$ as above,
\begin{equation}
|x-y|\ge\bigl|x-T_x(y)\bigr|
\end{equation}
where, we recall, $|x-y|$ is the Euclidean distance of~$x$ and~$y$.
\end{lemma}

\begin{proofsect}{Proof}
Letting $i$ and~$j$ be as above, we may assume, without loss of generality, that $x_i>0$ --- otherwise this may be achieved by reflecting all components of all vectors. Abbreviate $\tilde y:=T_x(y)$. If $i=j$, then we have $x=x_i\hate_i+x'$ and $y=y_i\hate_i+y'$ where $x',y'$ are orthogonal to~$\hate_i$. A calculation shows
\begin{equation}
|x-y|^2-|x-\tilde y|^2=(x_i-y_i)^2-(x_i-\tilde y_i)^2=2x_i(\tilde y_i-y_i)
\end{equation}
which is positive because $\tilde y_i=|y_i|\ge y_i$ (and $x_i>0$ by assumption).

The second case is $i\ne j$. Here we will write $x=x_i\hate_i+x_j\hate_j+x'$, $y=y_i\hate_i+y_j\hate_j+y'$ and $\tilde y=\tilde y_i\hate_i+\tilde y_j\hate_j+y'$ where $x'$ and $y'$ are orthogonal to $\hate_i$ and $\hate_j$. A calculation shows
\begin{equation}
\begin{aligned}
|x-y|^2-|x-\tilde y|^2&=
2x_i(\tilde y_i-y_i)+2x_j(\tilde y_j-y_j)
\\
&=2x_i\bigl(|y_j|-y_i\bigr)+2x_j\bigl(\sign(y_j)|y_i|-y_j\bigr)
\\
&=2x_i\bigl(|y_i|-y_i\bigr)+2\bigl(x_i-\sign(y_j)x_j\bigr)\bigl(|y_j|-|y_i|\bigr).
\end{aligned}
\end{equation}
Both terms are non-negative because $x_i>0$, $|y_i|\ge y_i$, $x_i\ge \sign(y_j)x_j$ and $|y_j|\ge |y_i|$.
\end{proofsect}

\begin{proofsect}{Proof of Lemma~\ref{lemma4.3}}
By the Assumption~\ref{ass}(2) we have $\tilde c|x-y|^{-s}\le K_{xy}\le \tilde c^{-1}|x-y|^{-s}$  for some $\tilde c\in(0,1)$ whenever $|x-y|$ is sufficiently large. So let us denote
\begin{equation}
\widetilde Q_i^-:=\sum_{(x,y)\in\PP_i^+}\frac{|\theta_x-\theta_y|^2}{|x-y|^s},
\end{equation}
where
\begin{equation}
\PP_i^+:=\bigl\{(x,y)\in\PP_i\colon x_1\ge\max_k|x_k|,\,T_x(y)=y\bigr\}.
\end{equation}
Note that $T_x$ maps $\PP_i$ into $\PP_i^+$ in (at most) $2d$-to-one fashion and (being derived from $\ell^\infty$-distance) it preserves the deformation angles, i.e., $\theta_y=\theta_{T_x(y)}$.
An argument as in the proof of Lemma~\ref{L:Aplus} allows us to replace $K_{xy}+K_{yx}$ by $|x-y|^{-s}$ and estimate the finite-distance corrections by $\tilde c'a^{-1}L^{d-1}$. For a lower bound on the resulting sum we restrict the summation over $\PP_i$ to $\PP_i^+$; for an upper bound we first invoke Lemma~\ref{lemma4.4} to dominate the sum over $\PP_i$ by $2d$-times the sum over $\PP_i^+$. This yields 
\begin{equation}
2\tilde c\widetilde Q_i^--\tilde c'a^{-1}L^{d-1}\le Q_i^-\le(2d)2\tilde c^{-1}\widetilde Q_i^-+\tilde c'a^{-1}L^{d-1}.
\end{equation}
It will thus suffice to study the asymptotic of $\widetilde Q_i^-$ alone.

On $\PP_i^+$, both $x_1,y_1$ are positive with $y_1>L-a$. So we may define
\begin{equation}
u:=|x_1-L+a|\quad\text{and}\quad t:=y_1-(L-a).
\end{equation}
Notice that then
\begin{equation}
\frac{|\theta_y-\theta_x|^2}{\pi^2}=
\begin{cases}
(\ffrac ta)^2,\qquad&\text{if }(x,y)\in\PP_1^+,
\\
1,\qquad&\text{if }(x,y)\in\PP_2^+,
\\
\bigl(\frac{t-u}a\bigr)^2,\qquad&\text{if }(x,y)\in\PP_3^+,
\\
(\frac{a-u}a)^2,\qquad&\text{if }(x,y)\in\PP_4^+.
\end{cases}
\end{equation}
We will also denote by~$z$ the projection of $y-x$ to the direction orthogonal to~$\hate_1$.

The bounds on $\widetilde Q_i^-$ are now generally carried out as follows. We fix the component of~$x$ orthogonal to~$\hate_1$ to some $d-1$ dimensional vector $\hat x$ and derive uniform bounds on the sum over $t$, $u$ and~$z$. Then we sum over the number of admissible $\hat x$'s --- this number will inevitably be of order $L^{d-1}$. For $\widetilde Q_1^-$ and~$\widetilde Q_4^-$ an inspection of \eqref{Eq1}, resp., \eqref{Eq4} shows
\begin{equation}
\widetilde Q_i^-\le \pi^2(2L+1)^{d-1}q_i,\qquad i=1,4,
\end{equation}
where $(2L+1)^{d-1}$ is an upper bound on the number of~$x$'s contributing for a given~$t$ and where we performed a change of variables $u\mapsto a-u$ to get the stated form of $q_4$.
The same method will produce a corresponding upper bound also in the cases $\widetilde Q_2^-$ and $\widetilde Q_3^-$ with~$q_{i,\infty}$ on the right-hand side. For the lower bounds on $\widetilde Q_2^-$ and $\widetilde Q_3^-$ we instead restrict~$x$ further so that $|\hat x|_\infty\le L/2$ (here is where we use $a\le L/2$). Then $z$ can be summed freely as long as $z\in R_L$. This yields
\begin{equation}
\pi^2L^{d-1}q_{i,L}\le \widetilde Q_i^-\le \pi^2(2L+1)^{d-1}q_{i,\infty},\qquad i=2,3.
\end{equation}
Combining the above observations and invoking \eqref{E:4.8}, the claim follows.
\end{proofsect}

We are now ready to finish the proof of \eqref{E:I-bounds}:

\begin{proofsect}{Proof of Lemma~\ref{L:Aminus}}
Lemma~\ref{lemma4.3} reduces the claim to finding proper leading-order expressions for the quantities $q_1,q_{2,N},q_{3,N},q_4$ above. To keep the expressions simple, let us agree to write $f\asymp g$ if the ratios $f/g$ and $g/f$ are bounded by universal constants depending only on~$d$ and~$s$, uniformly in~$a,L$ subject to the bounds $c^{-1}<a<cL$ for some given small~$c\in(0,1)$.

We begin by noting that, for any integer~$m$ with $1\le m\le 2L$ and $s>d$, we have
\begin{equation}
\label{E:4.19}
\sum_{z\in R_L}\frac1{(m^2+|z|^2)^{s/2}}\asymp \sum_{z\in\Z^{d-1}}\frac1{(m^2+|z|^2)^{s/2}}\asymp m^{d-1-s}.
\end{equation}
This immediately implies that $q_{2,L}\asymp q_{2,\infty}$ and $q_{3,L}\asymp q_{3,\infty}$ and so we can treat both terms on the same footing. As for~$q_1$, \eqref{E:4.19} permits us to write
\begin{equation}
q_1\asymp a^{-2}\sum_{u=0}^L\sum_{t=1}^a t^2(u+t)^{d-1-s}\asymp a^{d+1-s},
\end{equation}
where we first summed over~$u$ assuming $s>d$ and then summed over~$t$ employing $d+2-s\ge0$. Similarly we get
\begin{equation}
q_{2,\infty}\asymp\sum_{u=0}^L\sum_{t> a}(u+t)^{d-1-s}\asymp
\begin{cases}
a^{d+1-s},\qquad&\text{if }s>d+1,
\\
\log\bigl(\frac{L+a}a\bigr),\qquad&\text{if }s=d+1,
\\
(L+a)^{d+1-s},\qquad&\text{if }s<d+1.
\end{cases}
\end{equation}
Here we first summed over~$t$ and then distinguished the three possibilities depending on whether the remaining sum is divergent, logarithmically divergent and convergent.

For the remaining two terms we get the following: In light of \eqref{E:4.19} and the fact that absolute constants do not matter, we get
\begin{equation}
q_{3,\infty}\asymp a^{-2}\sum_{u=0}^a\sum_{t=u+1}^a(t-u)^{d+1-s}\asymp
\begin{cases}
a^{d+1-s},\qquad&\text{if }s<d+2,
\\
a^{-1}\log a,\qquad&\text{if }s=d+2,
\end{cases}
\end{equation}
where we only paid attention to the values of~$s$ with $d<s\le d+2$. Finally we get
\begin{equation}
q_4\asymp a^{-2}\sum_{u=0}^a\sum_{t\ge0}u^2(u+t)^{d-s-1}\asymp a^{d+1-s},
\end{equation}
where we employed that $s<d+3$.

It is now straightforward to check that, for $1\ll a\ll L$, the dominant term for $s\in(d,d+1]$ is $q_{2,\infty}\asymp q_{2,L}$ while for $s\in(d+1,d+2]$ the dominant term is $q_{3,\infty}\asymp q_{3,L}$. Combining this with the conclusions of Lemma~\ref{lemma4.3}, the claim follows.
\end{proofsect}

\section{Expected energy defect}
\label{sec5}\noindent
Our final task is to establish Proposition~\ref{P:Delta-mstar}. Fix~$L,a\in\D$ with~$L>a$ and recall the notation $\cmss R^\pm$ for the inhomogeneous rotations from \eqref{E:Rpm}. For any~$x,y$ let
\begin{equation}
\Delta_{xy}(\sigma):=2\sigma_x\cdot\sigma_y-(\cmss R^+\sigma)_x\cdot(\cmss R^+\sigma)_y-(\cmss R^-\sigma)_x\cdot(\cmss R^-\sigma)_y
\end{equation}
denote the term corresponding to these vertices from the long-range part of the energy defect~$\Delta_{L,a}$.
Abbreviate
\begin{equation}
\widetilde K_{xy}:=4\sin^2\biggl(\frac{\theta_x-\theta_y}2\biggr)K_{xy}
\end{equation}
and let $\cmss P_{12}$ denote the orthogonal projection of~$\R^n$ onto the linear span of $\hate_1,\hate_2$.
We begin with a variation on Lemma~4.4 from~\cite{BCK}:

\begin{lemma}
\label{L-smoothing}
Suppose Assumption~\ref{ass}(2) holds. For an integer~$\ell\ge1$, let $V_1$ and~$V_2$ be two disjoint translates of~$\Lambda_\ell$.
For each~$\epsilon>0$ there is~$\delta>0$ such that if $\dist(V_1,V_2)\ge \ell/\delta$ and $\ell/a<\delta$, then for all~$\sigma\in\Omega$,
\begin{equation}
\biggl|\,\sum_{x\in V_1}\sum_{y\in V_2}
K_{xy}\Delta_{xy}(\sigma)-m_1(\sigma)\cdot m_2(\sigma)\sum_{x\in V_1}\sum_{y\in V_2}\widetilde K_{xy}\biggr|
\le\epsilon \sum_{x\in V_1}\sum_{y\in V_2}\widetilde K_{xy},
\end{equation}
where $m_i(\sigma):=|\Lambda_\ell|^{-1}\sum_{x\in V_i}\cmss P_{12} \sigma_x$ is the $\cmss P_{12}$-projection of the spin average in~$V_i$.
\end{lemma}

\begin{proofsect}{Proof}
As is easy to check from \eqref{E:Rtheta}, we have
\begin{equation}
\label{E:5.4trick}
\Delta_{xy}(\sigma)=4\sin^2\biggl(\frac{\theta_x-\theta_y}2\biggr)\,(\sigma_x\cdot\cmss  P_{12}\sigma_y)
\end{equation}
and so $K_{xy}\Delta_{xy}(\sigma)=\widetilde K_{xy}(\sigma_x\cdot\cmss  P_{12}\sigma_y)$.
Now pick~$x_0\in V_1$ and $y_0\in V_2$. Assumption~\ref{ass}(2) ensures that, for each~$\epsilon>0$ there is~$\delta>0$ such that if
\begin{equation}
|x-y|\ge\delta^{-1}\max\{|x-x_0|,|y-y_0|\},
\end{equation}
then
\begin{equation}
\bigl|K_{xy}-K_{x_0y_0}\bigr|\le \epsilon K_{x_0y_0},\qquad x\in V_1,\,y\in V_2.
\end{equation}
Since $|\theta_x-\theta_{x_0}|\le\frac{\pi}{a}|x-x_0|\le\pi{\ell}/{a}<\pi\delta$, a similar bound holds also for $\widetilde K_{xy}$. The claim is now proved as in~\cite[Lemma~2.2]{BCK}.
\end{proofsect}

\begin{proofsect}{Proof of Proposition~\ref{P:Delta-mstar}}
Consider a translation-invariant, ergodic Gibbs measure~$\mu$ satisfying \eqref{E:2.7}. Recall the notation $m_i(\sigma)$ from Lemma~\ref{L-smoothing}. For any~$\epsilon>0$, let
\begin{equation}
\EE_\ell:=\biggl\{\sigma\colon\Bigl|\,\sum_{x\in\Lambda_\ell}\sigma_x-m_\star\hate_1|\Lambda_\ell|\Bigr|<\epsilon|\Lambda_\ell|\biggr\}.
\end{equation}
By the Spatial Ergodic Theorem, there exists $\ell_0=\ell_0(\epsilon)$ such that for $\ell\ge\ell_0$ we have
$\mu(\EE_\ell)\ge1-\epsilon$. Thus, if $\ell\ge\ell_0$ and $V_1$ and~$V_2$ are disjoint translates of~$\Lambda_\ell$, then
\begin{equation}
\Bigl|E_\mu\bigl(m_1(\sigma)\cdot m_2(\sigma)\bigr)-m_\star^2\Bigr|<5\epsilon.
\end{equation}
Assuming that $\dist(V_1,V_2)\ge\ell/\delta$ and $\ell/a<\delta$, Lemma~\ref{L-smoothing} shows
\begin{equation}
\label{E:5.9}
E_\mu \biggl(\,\sum_{x\in V_1}\sum_{y\in V_2}
K_{xy}\Delta_{xy}(\sigma)\biggr)\ge (m_\star^2-6\epsilon)\sum_{x\in V_1}\sum_{y\in V_2}
\widetilde K_{xy}.
\end{equation}
Now consider a fixed partition of~$\Z^d$ into blocks of side~$\ell$. Summing \eqref{E:5.9} over the blocks in the partition, and applying \eqref{E:5.4trick} one more time we get
\begin{equation}
\label{E:Big}
\begin{aligned}
E_\mu\bigl(\Delta_{L,a}\bigr)\ge\,\,(m_\star^2-6\epsilon)\!\!\sum_{\begin{subarray}{c}
x,y\\|x-y|\ge2\ell/\delta
\end{subarray}}\!\!
\widetilde K_{xy}\,\,
-\!\!\sum_{\begin{subarray}{c}
x,y\\|x-y|\le2\ell/\delta
\end{subarray}}
\widetilde K_{xy}
\\*[2mm]
\ge (m_\star^2-6\epsilon)\sum_{x,y\colon x\ne y}
\widetilde K_{xy}
-\,2\!\!\!\sum_{\begin{subarray}{c}
x,y\\|x-y|\le2\ell/\delta
\end{subarray}}
\widetilde K_{xy},
\end{aligned}
\end{equation}
where we used $0<m_\star^2-6\epsilon<1$.
It remains to bound the terms on the right-hand side.

Using $\widetilde K_{xy}\ge (4/\pi^2)|\theta_x-\theta_y|^2K_{xy}$ and Lemma~\ref{L:Aplus}, the first sum is at least a constant times $\II_{L,a}$. For the second sum we note that for all contributing~$x,y$ we have
\begin{equation}
\widetilde K_{xy} \leq K_{xy}|\theta_x-\theta_y|^2\le c_1 \left(\frac{\ell}{\delta a}\right)^2,
\end{equation}
where $c_1:=\sup K_{0,x}$. Moreover, $\widetilde K_{xy}$ is zero unless at least one of~$x$ and~$y$ lies in the annulus $\Lambda_L\setminus\Lambda_{L-a}$. This implies
\begin{equation}
\,\,\!\!\sum_{\begin{subarray}{c}
x,y\\|x-y|\le2\ell/\delta
\end{subarray}}
\widetilde K_{xy} \le c_2 L^{d-1} \frac{\ell^{d+2}}{\delta^{d+2} a}
\end{equation}
for some~$c_2$ proportional to~$c_1$ above. If $a$ is so large that one can find $\ell\ll\delta [\epsilon\log a]^{\frac1{d+2}}$ with $\ell\ge \ell_0$, then the right hand side is at most $\epsilon L^{d-1} a^{-1}\log a$. As this is much smaller than $\II_{L,a}$ for all $s\in(d,d+2]$, the claim follows.
\end{proofsect}

\section*{Acknowledgments}
\noindent
We wish to thank A.\ van Enter for suggestions over an earlier version of this paper and for proposing one of the questions in Remarks~\ref{R}(4). The research of M.B.\ was partially supported by the NSF grant DMS-0949250.  The research of N.C. was supported in part by a Marilyn and Michael Winer Fellowship and by the Binational Science Foundation Grants BSF-2008421 and BSF-2006477. We express our gratitude to anonymous referees for interesting suggestions on the first version of this paper and for observant remarks that made us realize an omission of some important cases in one of the essential calculations.

\end{document}